\documentclass[conference]{IEEEtran}
\usepackage{cite}
\usepackage{amsmath,amssymb,amsfonts}
\usepackage{algpseudocode,algorithm}
\usepackage{graphicx}
\usepackage{textcomp}
\usepackage{xcolor}
\usepackage[colorlinks,allcolors=blue]{hyperref}
\begin{document}

\title{BLANT: Basic Local Alignment of Network Topology, Part 1: Seeding local alignments with unambiguous 8-node graphlets}

\author{\IEEEauthorblockN{Patrick Wang$^\dagger$}
\IEEEauthorblockA{\textit{Department of Computer Science} \\
\textit{University of California, Irvine}\\
Irvine, United States \\
wangph1@uci.edu\\
$^\dagger$Equal contributor}
\and
\IEEEauthorblockN{Henry Ye$^\dagger$}
\IEEEauthorblockA{\textit{Department of Computer Science} \\
\textit{University of California, Irvine}\\
Irvine, United States \\
hanweny@uci.edu \\
$^\dagger$Equal contributor}
\and
\IEEEauthorblockN{Wayne Hayes$^*$}
\IEEEauthorblockA{\textit{Department of Computer Science} \\
\textit{University of California, Irvine}\\
Irvine, United States \\
whayes@uci.edu \\
$^*$Corresponding author}
}

\maketitle

\begin{abstract}
BLAST is a standard tool in bioinformatics for creating local sequence alignments using a ``seed-and-extend'' approach. Here we introduce an analogous seed-and-extend algorithm that produces local {\em network} alignments: BLANT, for {\em Basic Local Alignment of Network Topology}. This paper introduces {\em BLANT-seed}: given an input graph, BLANT-seed uses network topology alone to create a limited, high-specificity index of {\em k}-node induced subgraphs called {\em k}-graphlets (analogous to BLASTS's {\em k}-mers). The index is constructed so that, if significant common network topology exists between two graphs, their indexes are likely to overlap. BLANT-seed then queries the indexes of two networks to generate a list of common {\em k}-graphlets which, when paired, form a {\em seed pair}. Our companion paper (submitted elsewhere) describes {\em BLANT-extend}, which ``grows'' these seeds to larger local alignments, again using only topological information.
\end{abstract}


\section{Introduction}
Network alignment is the task of finding topologically similar network regions over a database of networks with applications in a wide variety of fields \cite{emmert2016fifty}. These include bioinformatics \cite{clark2014comparison}, linguistics \cite{Dehmer2011}, neuroscience \cite{Sommerfeld1994}, social networks \cite{metaDiagramAlignment}, and image recognition \cite{HSIEH2008401}. Being a generalization of subgraph isomorphism, it is NP-hard \cite{GareyJohnson}, so many heuristic algorithms exist to approximately solve it. In this paper we consider only the case of aligning \textit{two} networks.

Network alignment can be either local or global \cite{meng2016local}. Global network alignment aims to find a 1-to-1 mapping from the nodes of the smaller network to the nodes of the larger one. In contrast, local network alignment aims to find highly similar regions across the networks, and allows each node to appear in more than one such local alignment.

Both of these problems have been studied extensively, and a key characteristic of any network alignment algorithm is whether it uses information beyond the graph's topology. On one end of the spectrum, the alignment algorithm may heavily rely on domain-specific node annotations like usernames in a social network \cite{LIU2018318} or nucleotide sequences in a biological network \cite{alignMCL}. On the other end of the spectrum, there are alignment algorithms \cite{MamanoHayesSANA,regal2018,dana2019} which align solely using topological information. These algorithms remove the significant cost of gathering domain-specific knowledge for each node and can generalize to networks in any domain. Additionally, such algorithms may discover unique insights hidden in the topology of the graph itself \cite{wang2022sana,WeBeat}.

However, the topology-only algorithms listed above are all global alignment algorithms. To the best of our knowledge, there do not currently exist any local alignment algorithms which use only topological information\footnote{Although our earlier work, GRAAL \cite{GRAAL}, is topology-only and also used a seed-and-extend approach, we do not consider it here because, based on letting it run initially for several days on the IID networks---see below---its own estimate of its runtime was upwards of 2200 hours, and we chose not to wait 3 months for the output.}. BLANT-seed + BLANT-extend seeks to fill this gap in the state-of-the-art.

\begin{figure}
    \centering
    \includegraphics[width=0.4\textwidth]{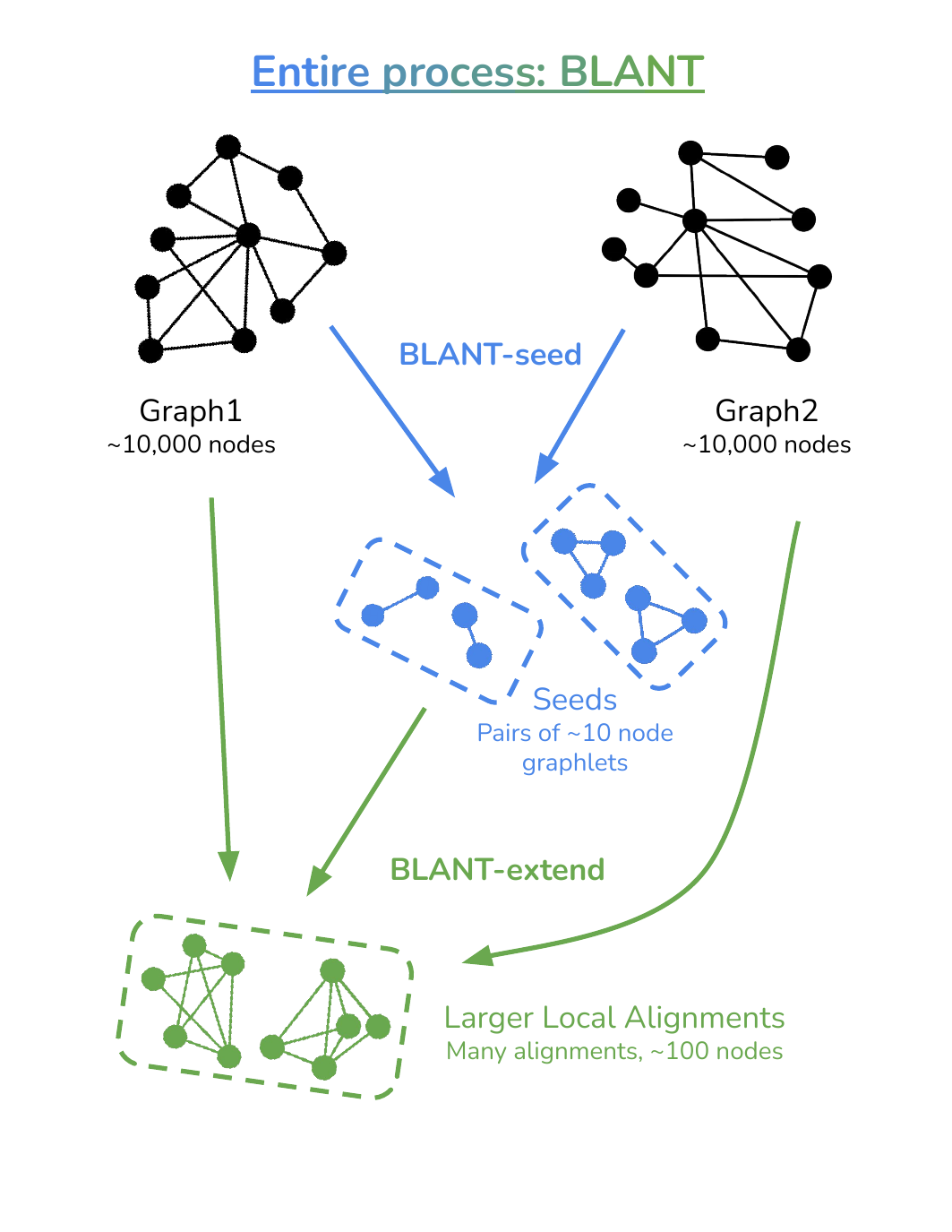}
    \caption{ BLANT-seed produces small, perfect local alignments we will also call \textbf{seeds}, while BLANT-extend extends these into larger local alignments that are not required to be perfect. Collectively, the BLANT-seed + BLANT-extend pipeline is referred to as ``BLANT''. }
    \label{fig:blant_both}
\end{figure}

BLANT-seed focuses heavily on $k$-graphlets, which are induced subgraphs of $k$ nodes. A pair of isomorphic graphlets forms a seed, and BLANT-seed is capable of outputting seeds of up to $k=15$ nodes. A few other seed identification methods exist (such as ODV similarity \cite{Tijana2008}), but BLANT-seed is unique in its ability to discover seeds of more than one node pair. This is significant because BLANT-extend \cite{BLANT-extend} demonstrates that graphlet-seeds are significantly more effective than simple node pairs when expanding into large, high-quality alignments. In fact, BLANT-extend was able to extend our graphlet-seeds into two 100\% isomorphic subgraphs of 280 nodes each between two networks with about 15,000 nodes and 300,000 edges. Fig.~\ref{fig:blant_both} shows a high-level overview of the BLANT-seed + BLANT-extend pipeline.


BLANT started out \cite{blanttool,hayesblant} as a random sampling algorithm to approximate the global distribution of graphlets in a large network, allowing a high-quality fast approximation to the far more expensive exhaustive enumeration offered by codes like ORCA \cite{ORCA} and Jessy \cite{Melckenbeeck1EtAl2017}. We will refer to this sampling algorithm as BLANT-sample. After significant effort spent experimenting with BLANT-sample's randomly sampled graphlets, we came to a crucial realization: random samples are exceedingly unlikely to find any graphlets from similar regions of the graph due to the exponential number of graphlets, and are thus exceedingly unlikely to discover good local alignments.

BLANT-seed, however, uses an innovative and counter-intuitive \textit{deterministic} approach of creating an index on each network. The indexes are then mined for topologically identical graphlets across networks, which form {\it seeds}. Determinism is key, because if the networks \textit{actually have similarity}, then running the same deterministic algorithm on each of them is more likely to result in sampling similar slices of the exponential search space, creating two indexes which are likely to lead to high quality alignments.


Our main contributions are the following:
\begin{enumerate}
    \item Collectively, BLANT-seed and BLANT-extend constitute the first local aligner which produces excellent results using topology alone, and we accomplish this on large and realistic networks.
    \item We develop the first algorithm which is capable of identifying seeds of more than one node. Specifically, BLANT-seed can find seeds of up to $k=15$ nodes with \textit{perfect} topological similarity and high functional similarity.
    \item We demonstrate the counter-intuitive power of \textit{determinism} in effectively indexing large network databases.
\end{enumerate}

\section{Background}

Our algorithm name---BLANT---is a clear ripoff of {\it BLAST}---the ubiquitous {\it Basic Local Alignment Search Tool} \cite{Altschul90}. This homage is justified, since both BLAST and BLANT are seed-and-extend approaches with many principled similarities. To see the analogy, we first review how BLAST works.

BLAST is an algorithm for quickly finding local alignments in genomic (or proteomic) sequences. BLAST works by first creating a comprehensive index of all $k$-letter sequences called ``$k$-mers'' that appear in the corpus of the sequence to be searched and/or aligned. By storing every $k$-mer and its location, BLAST can ``line up'' the regions around two identical $k$-mers, and then check to see if this local alignment ``extends'' further beyond the $k$-mers. Below we show a hypothetical alignment between two distant regions of sequence, both of which contain the boldfaced $k$-mer (isolated for clarity):

{\tt\small\centering
\vspace{1mm}
\noindent ACTAGAT{\it C}CAC{\it C}TCTAG {\bf GAGACCGT} GTTCTTCA{\it G}AGGTG\\
\noindent ACTAGAT{\it A}CAC{\it G}TCTAG {\bf GAGACCGT} GTTCTTCA{\it T}AGGTG\\
\vspace{1mm}
}

In the case above, even though the sequences contain minor differences (highlighted with italics), the boldfaced identical $k$-mers form a {\it seed} that can be used to attempt {\em extending} the match in both directions beyond the endpoints of the common $k$-mer. BLAST is extremely fast at performing the above operations, which is the reason BLAST has become the near-ubiquitous tool for aligning sequences that contain billions of letters. BLAST automatically chooses the appropriate value of $k$ to create $k$-mer seeds in a particular search and alignment task, and uses a sophisticated extend algorithm to create full seed-and-extend local alignments.

BLANT uses an analogous seed-and-extend approach but for networks: given an undirected network $G$ and a value of $k$, BLANT-seed (the algorithm in this paper) samples connected, induced $k$-node subgraphs called $k$-graphlets \cite{Przulj2004Graphlets}. Since the number of $k$-graphlets in a graph of $n$ nodes is exponential in both $k$ and $n$, BLANT-seed cannot exhaustively enumerate all $k$-graphlets, but instead deterministically samples them. If one is interested in the overall distribution of graphlets across a network, BLANT-sample \cite{blanttool,hayesblant} can randomly sample millions of graphlets per second to estimate this distribution. While this is useful, BLANT-seed takes the opposite approach: a {\em deterministic} expansion that, with care, produces graphlet seeds appropriate for expansion by BLANT-extend \cite{BLANT-extend}.

Starting with a given seed of two identical $k$-graphlets, the algorithm in our companion paper, BLANT-extend \cite{BLANT-extend}, attempts to extend the local alignment beyond the boundaries of the seed. While the $k$-graphlets of the seed must be identical---just like BLAST's $k$-mers---BLANT-extend allows for non-exact topological matching as it expands the local alignment beyond the seed. 


\subsection{Graphlets, Orbits, and Ambiguity}
\begin{figure}
    \centering
    \includegraphics[width=0.47\textwidth]{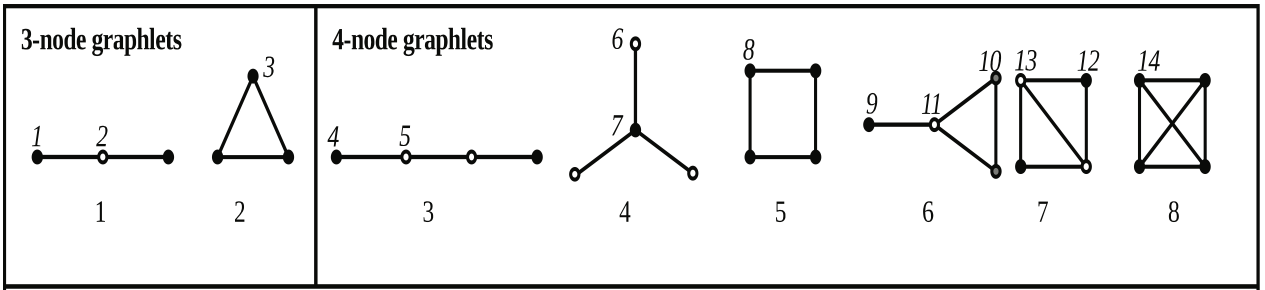}
    \caption{ All graphlets of sizes k = 3 and 4 nodes, and their automorphism orbits; within each graphlet, nodes of equal shading are in the same orbit. This figure is taken from \cite{Melckenbeeck1EtAl2016}. (Note that BLANT uses different, automatically-generated IDs than \cite{Melckenbeeck1EtAl2016}.) }
    \label{fig:graphlets}
\end{figure}

Graphlets are defined as ``small'' induced subgraphs of a network; they were originally exhaustively enumerated up to $k=5$ nodes \cite{Przulj2004Graphlets}, then $k=6$ nodes \cite{Melckenbeeck1EtAl2016,melckenbeeck2017efficiently}, though BLANT now exhaustively enumerates all graphlets with up to $k=8$ nodes. Fig.~\ref{fig:graphlets} shows the complete list of unique $k$-graphlets for $k=3,4$. BLANT automates the process up to $k=8$, in which there are $n_k=12,346$ unique graphlets containing a total of 79,264 orbits. Each $k$-graphlet is assigned a unique {\it graphlet ID} from $0$ to $n_k-1$ inclusive.

An orbit is set of nodes that are topologically identical within a graphlet; {\it ie.,} they can be swapped without changing the graphlet. In Fig.~\ref{fig:graphlets}, nodes are shaded based on their orbit. For example, the first 4-node graphlet in Fig.~\ref{fig:graphlets} is the path of length 3, which has two orbits: the two middle nodes together form one orbit, while the two end-nodes form another.

\begin{figure}
    \centering
    \includegraphics[width=0.49\textwidth]{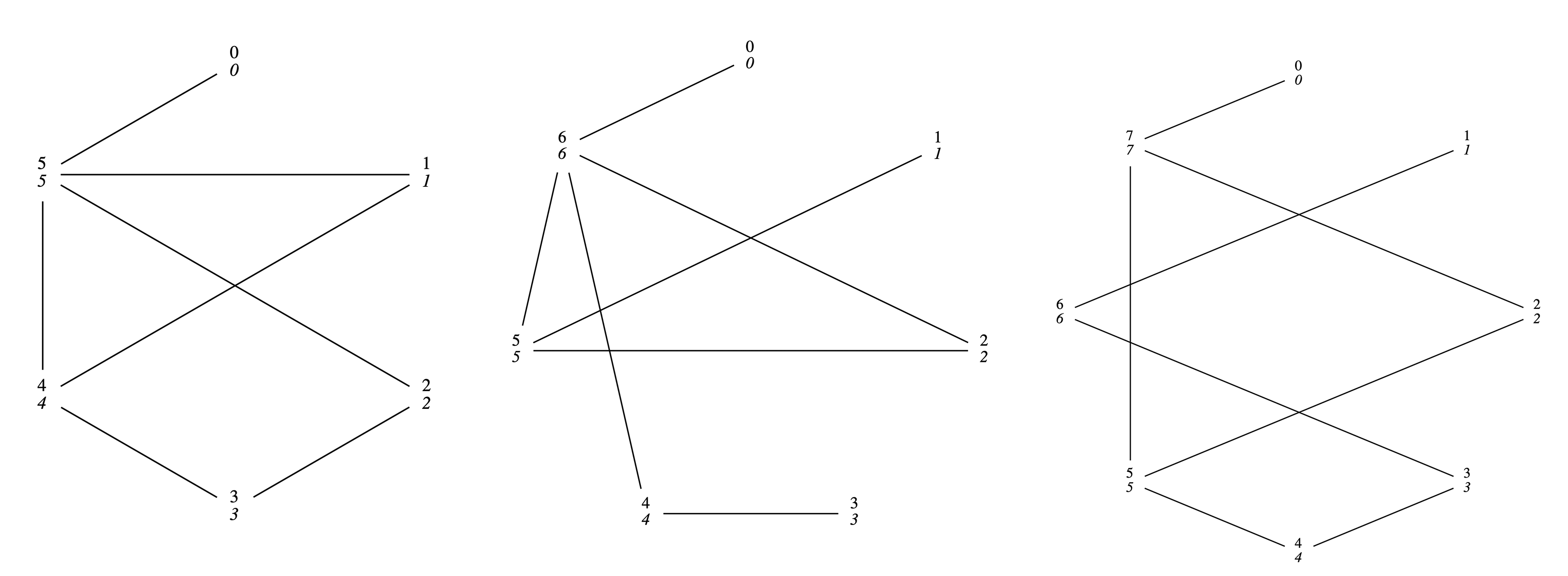}
    \caption{Examples of unambiguous graphlets on $k=6,7,$ and 8 nodes. Each node is represented by an integer from 0 to $k-1$, and its orbits are labeled in italics. As we can see, in each case each orbit contains only one node, making each of these graphlets unambiguous. }
    \label{fig:unambig_graphlets}
\end{figure}

When it comes time to extend a graphlet-based seed, multiple nodes with the same orbit are problematic as they can be swapped without changing the graphlet. This introduces an ambiguity that would severely complicate the process of extending the seed. To avoid having BLANT-extend deal with the resulting combinatorial explosion, we eliminate all such graphlets at the BLANT-seed stage. We define an unambiguous graphlet as a graphlet in which every orbit has only one node. Unfortunately, the traditional set of graphlets containing at most $k=5$ nodes have no unambiguous graphlets; the smallest value of $k$ for which unambiguous graphlets exist is $k=6$, and Fig.~\ref{fig:unambig_graphlets} provides a few examples.

\subsection{Alignments, and Types of Alignment Similarity}
\label{sec:eval_metrics}
We define an alignment as a 1-to-1 mapping from nodes in one network to nodes in another network. The number of nodes being mapped can range anywhere from one to the number of nodes in the smaller network.

Topological similarity refers to how close the aligned (sub)graphs are to being isomorphic; common measures include EC \cite{GRAAL} and $S^3$ \cite{MAGNA}, both of which count the number of aligned edges as a fraction of some maximum. {\em Functional} similarity refers to what the nodes and edges {\em are}, or what they {\em do}, in the real world. In the most extreme case of high functional similarity, two nodes could be considered ``correctly aligned'' if they are virtually {\em identical} in the appropriate context. For example, in social networks, two nodes in different social networks are identical if they refer to the same person; in Protein-Protein Interaction (PPI) networks, two proteins are considered identical if they are orthologs (related by a common ancestor). {\it Node Correctness} (NC) is commonly used as a stringent measure of post-hoc alignment quality: it is the fraction of output aligned node pairs that are correctly aligned.

The reader should note the difference between ``information'' and ``similarity''. Our algorithm uses only topological information but is evaluated based on topological and functional similarity.

\section{Deterministic Index Creation Algorithm}
\begin{figure}
    \centering
    \includegraphics[width=0.47\textwidth]{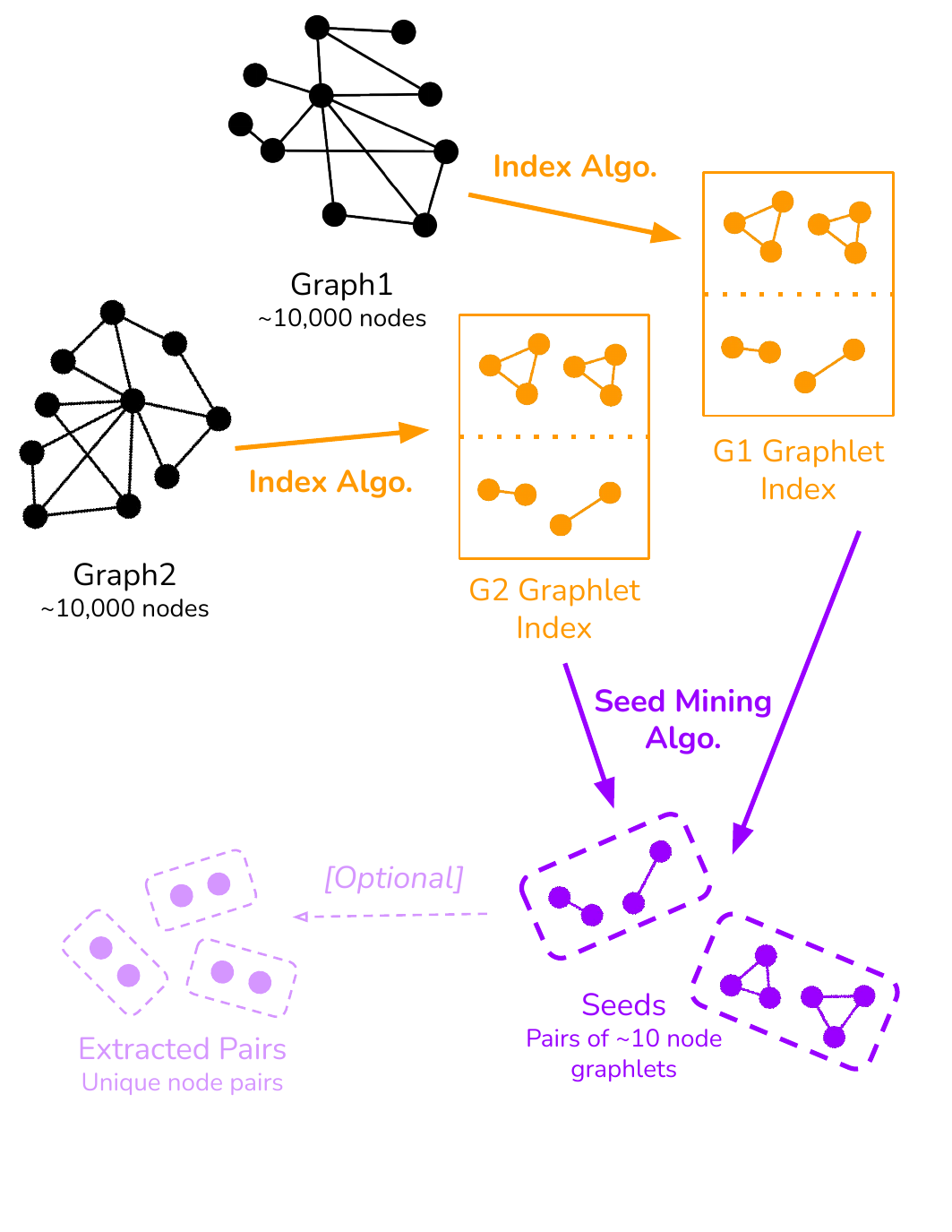}
    \caption{ The overall process for BLANT-seed. First, an index is generated deterministically for the two networks we wish to align. Then, we use a seed mining algorithm to query both indexes and output high quality seeds. Finally, we can optionally extract unique node pairs out of our graphlet seeds. }
    \label{fig:blant_seed_pipeline}
\end{figure}

\noindent There are two steps: indexing, and seed creation. This section is about indexing, and is depicted in orange in Fig.~\ref{fig:blant_seed_pipeline}.

\begin{algorithm}
\caption{Index Creation Algorithm}
\label{alg:index}
\begin{algorithmic}
    \Function{CreateIndex}{Graph $G$, $k$, $D$, $f$}
        \State sorted = $G$'s nodes sorted by $f$ descending
        \For{$u$ in sorted}
            \State index += GetNodeEntries($G$, $k$, $D$, $f$, [$u$])
            \State // Note: [$u$] above is $V_0$ (the root node) below
        \EndFor
        \State\Return{index}
    \EndFunction{}
    \State
    \Function{GetNodeEntries}{$G$, $k$, $D$, $f$, $V$[]}
        \If{$|V|=k$}
            \If{graphletID of $V$ is unambiguous}
                \State augId = graphletID + orbit($V_0$)
                \State\Return{IndexEntry(augId, $V$)}
            \EndIf{}
        \Else{}
            \State $V_{exp}$ = GetExpandNeighbors($G$, $V$, $D$, $f$)
            \For{$u$ in $V_{exp}$}
                \State $V=V\cup \{u\}$
                \State entries += GetNodeEntries($G$, $k$, $D$, $f$, $V$)
                \State $V=V-\{u\}$
            \EndFor
            \State\Return{entries}
        \EndIf{}
    \EndFunction
    \State
    \Function{GetExpandNeighbors}{$G$, $V$, $D$, $f$}
        \State neighs = \{all neighbors of nodes in $V$\}
        \State uniqueValues = \{unique $f$ values in neighs\}
        \State expandValues = \{top $D$ values in uniqueValues\}
        \State $V_{exp}$ = \{subset of neighs with $f$ values in expandValues\}
        \State\Return{$V_{exp}$}
    \EndFunction{}
\end{algorithmic}
\end{algorithm}

Index creation (Algorithm \ref{alg:index}) takes one network as input and returns an index with slightly augmented (cf. \S\ref{sec:bno}) graphlet IDs as keys. Each graphet ID maps to a limited set of graphlets of that ID found in the network with our deterministic algorithm.

The basic idea of this algorithm is to recursively build graphlets by deterministically expanding from some root node. To ensure that every node has a chance to be in the index, this expansion is performed $n$ times, once with each node in the network serving as the root. The algorithm builds graphlets in a DFS-like fashion, creating a list of nodes it will search in each recursive step. It iterates through the list of nodes, adding each one individually to its working set of nodes and recursing from there before backtracking and adding the next node to the working set. In order to deterministically create this list, the algorithm selects neighboring nodes with the highest values according to a deterministic heuristic function.

The graphlet IDs, orbit IDs, and (un)ambiguous determination are all extracted from standard output files from the current version of BLANT-sample \cite{blanttool,hayesblant}, which is itself based on \cite{hasan2017graphettes}.

\subsection{Heuristic Function}
The index creation algorithm relies on a heuristic function, $f(v)$, which deterministically assigns a numerical value to each node $v$ in the network. This value must be topologically meaningful, or else the deterministic algorithm will not be able to exploit topological similarity between the networks. For now $f(v)$=degree$(v)$ works well enough; we leave more sophisticated orderings to future work.

The heuristic function does not need to be injective. If there are multiple nodes of the same heuristic value, the algorithm will expand to all such nodes in order to maintain determinism. However, it is important to not have too many nodes of the same heuristic value, as this may result in unacceptably large runtimes.

\subsection{Augmenting Graphlet ID with the Root Node's Orbit}
\label{sec:bno}
Augmenting the graphlet ID with the orbit of the root node is a simple but effective technique that improves our results with no downside. Instead of only outputting the graphlet ID for a given graphlet, we create an augmented string which consists of the graphlet ID and the numerical orbit of the root node we expanded from to create that graphlet. This comes into play in our seed mining algorithm, which creates alignments based on key equality.

\subsection{Time Complexity}
\label{sec:index_time_complexity}
The algorithm performs a limited, recursive DFS-like search starting from each of the $n$ nodes of the network. At each recursive step, we visit all nodes with the top $D$ values of $f$, visiting at most $M$ nodes. Since the recursion depth is at most $k-1$, the time complexity for a search starting at one node is $O(M^{k-1})$, giving the overall algorithm a time complexity of $O(nM^{k-1})$. Assuming that the heuristic function contains relatively few ties---which is true when we are looking at the high end of degrees---$M$ will be approximately equal to $D$. In \S\ref{sec:patching_parameters} we show $D=2$ runs in a reasonable amount of time with our datasets.

\section{Index-Powered Seed Mining Algorithm}
\noindent There are two steps: indexing, and seed creation. This section is about seed mining, and is depicted in purple in Fig.~\ref{fig:blant_seed_pipeline}.
\begin{algorithm}[t]
\caption{Seed Mining Algorithm}
\label{alg:seed}
\begin{algorithmic}
    \Function{FindSeeds}{File $F_1$, File $F_2$, $C$, $P$}
        \State // $F_1$ and $F_2$ are the output files of Alg. \ref{alg:index}
        \State $I$ = PatchIndex($F_1$, $C$, $P$)
        \State $J$ = PatchIndex($F_2$, $C$, $P$)
        \State $K$ = $I$.keys $\cup$ $J$.keys
        \For{$k$ in $K$}
            \If{$|I_k|=1$ \textbf{and} $|J_k|=1$}
                \State $S=S\cup \{(I_{k_0},J_{k_0})\}$
            \EndIf
        \EndFor
        \State\Return $S$
    \EndFunction{}
    \State
    \Function{PatchIndex}{$F$, $C$, $P$}
        \For{$g$ in $F$}
            \For{$h$ in nearest $P$ graphlets to $g$}
                \State $CC$ = \# of common nodes between $g$ and $h$
                \If{$CC\ge C$}
                    \State // Note: + below is string concatenation
                    \State $key$ = $id(g)$ + $id(h) + CC +$ extra edges
                    \State $p$ = $g$.nodes $\cup$ $h$.nodes
                    \State $I_{key}=I_{key}\cup \{p\}$
                \EndIf{}
            \EndFor
        \EndFor{}
        \State\Return $I$
    \EndFunction{}
    \State
    \Function{ExtractNodePairs}{Seeds $S$}
        \For{$n$, $m$ in $S$}
            \State increment vote$_{n_m}$
        \EndFor
        \For{$n$ in vote}
            \State fav$_n$ = highest count nodes in vote$_n$
        \EndFor
        \State\Return node pairs which favorite each other
    \EndFunction
\end{algorithmic}
\end{algorithm}

Algorithm \ref{alg:seed} takes in the indexes of any number of networks and mines a limited list of perfect $k$-graphlet alignments between them. Each local alignment is called a {\em seed} and consists of a set of identical $k$-graphlets that appear across the networks. Note that, though the seeding process can find local alignments between any set of networks, for simplicity in this paper we restrict ourselves to aligning two networks at a time.


If graphlet pairs are desired, the algorithm ends here. If node pairs are desired, a simple voting algorithm---``how many graphlet pairs claim that these two nodes are aligned?''---is used to extract them from the graphlet pairs.

\subsection{Volume and Accuracy of Seeds}
The {\it Volume} of a set $S$ of $k$-graphlet seed pairs is $|S|$.
Assuming we know {\it a priori} some set of ``correct'' aligned pairs (eg., orthologs in biological networks, or the same person across two social networks), then the {\it Accuracy} of $S$ refers to the number of $k$-graphlet seeds in which some or all $k$ of the aligned nodes are correct. Given a set $T$ of unique $k=1$ seed pairs ({\it ie.,} node pairs), its volume is $|T|$, and its accuracy is the percentage of node pairs that are correct (cf. \S \ref{sec:metrics}).

\subsection{Index Key Specificity}
\label{sec:specificity}
A crucial step in solving in any graph alignment algorithm is pruning the search space.
Index key specificity is based on the observation that larger alignments are more likely to be constructed if the graphlet IDs in the seeds are rare. Say the graphlet ID $g$ appears $n_1$ times in the first index and $n_2$ times in the second index. Assuming each of the nodes in a graphlet has a single counterpart (which is generally true), the maximum possible number of {\it correct} alignments is $\min(n_1,n_2)$. Without another method to disambiguate these graphlet pairs, we must output all $n_1n_2$ pairs, capping our perfect alignment rate at $\min(n_1,n_2)/(n_1n_2)$. We have found empirically that it is best to impose the most stringent constraint: $n_1=n_2=1$.

\subsection{Patching Graphlets}
\label{sec:patch}
As we only align graphlets with identical keys \textit{and} when both keys are unique in their index, the performance of the algorithm depends on the number of keys indexed per network, as well an on the uniqueness of $k$-graphlets in a network. Just as with BLAST, larger values of $k$ increase the chance of having unique $k$-mers, but there is a tradeoff. Large $k$-mers and $k$-graphlets as sequencing errors or missing edges are more likely to appear. While BLAST automatically determines a good choice for this trade-off, the optimal value of $k$ for graphlets in our application is not yet known. However, we have found empirically that ``patching'' overlapping graphlets in the index into larger graphlets improves our results.

Natively, BLANT-sample can only sample graphlets up to $k=8$ nodes. As shown in \autoref{tab:unambiguous}, there are only 3034 unambiguous graphlets on 8 nodes, which turns out not to be enough in practice. To solve this, we ``patch'' together graphlets from the index which have common nodes in order to create a ``patched index'' of larger graphlets that have a higher likelihood of being unique\footnote{Creating the ``patched index'' could be considered part of the index creation algorithm. However, because this process is so fast, we decided to compute the patched index on the fly. This also increases flexibility, as we can easily experiment with different patching parameters.}.

\begin{table}[htbp]
\caption{Number of Unambiguous Graphlets for \textit{k} Values}
\label{tab:unambiguous}
\begin{center}
\begin{tabular}{|c|c|c|c|c|c|c|}
    \hline
    \textbf{k} & 3 & 4 & 5 & 6 & 7 & 8 \\
    \hline
    \textbf{\#} & 0 & 0 & 0 & 26 & 214 & 3034 \\
    \hline
\end{tabular}
\end{center}
\end{table}

\begin{figure}
    \centering
    \includegraphics[width=0.4\textwidth]{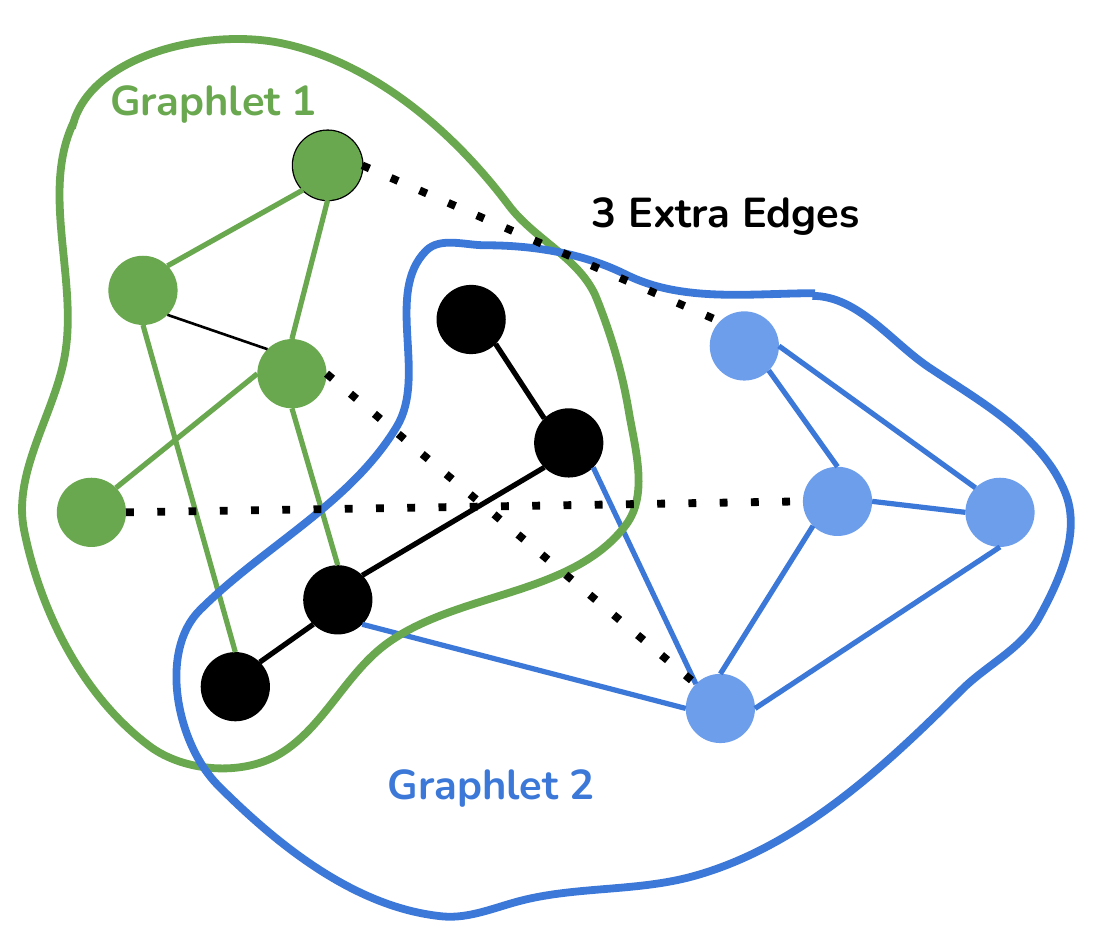}
    \caption{ A diagram of two 8-node graphlets with 4 nodes in common being patched together. The new patch ID is a string containing the graphlet ID of the first graphlet, the graphlet ID of the second graphlet, the nodes (technically orbits) which overlap, and the additional edges. }
    \label{fig:patch_diagram}
\end{figure}

Fig.~\ref{fig:patch_diagram} shows an example of two graphlets being patched together. After patching, a new ``patch ID'' must be created that encodes the shape of the larger graphlet. Although this ID is not completely unique (all ID's refer to a single shape but multiple ID's may refer to the same shape), we have found that it is unique enough so that collisions are rare. Additionally, we have found that canonizing this ID using a tool like NAUTY \cite{NAUTY} slightly decreases accuracy---a surprising result whose study we leave for future work.

\begin{figure}
    \centering
    \includegraphics[width=0.4\textwidth]{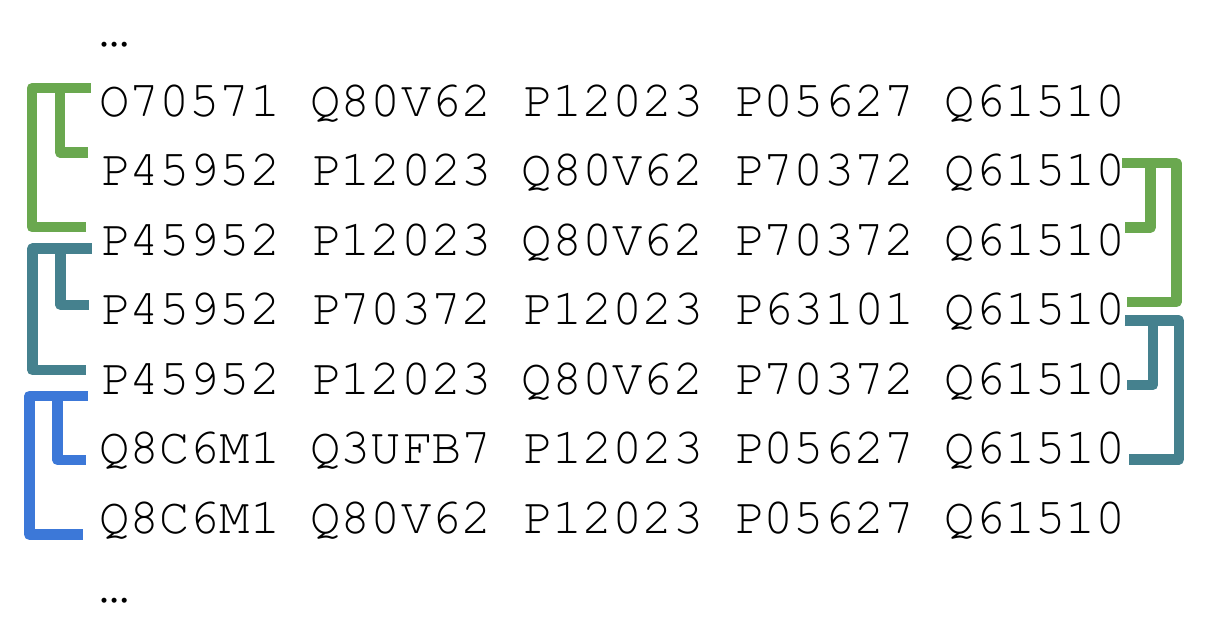}
    \caption{ A diagram showing which graphlets in the output file would be patched together when $P=2$. Each graphlet is compared with the two below it, and patched if they have at least $C$ nodes in common. } 
    \label{fig:prox}
\end{figure}

Conveniently, our index creation algorithm provides the perfect way to find graphlets to patch together: because the algorithm performs DFS, graphlets with many nodes in common will be close to each other in the output file. Thus, we simply go through each line of the output file and compare the graphlet with the $P$ lines below it (cf. Fig.~\ref{fig:prox}). In addition to saving time, this method also increases accuracy as we are only patching graphlets which were discovered close to each other in the deterministic run. This result is demonstrated in \S\ref{sec:patching_parameters}.

The second patching parameter, $C$, denotes the minimum number of nodes the graphlets must have in common to be patched. Smaller values of $C$ lead to larger graphlets, but the trade-off is that the patched graphlet will be less related to the graphlets in the index. The effects of $C$ are investigated in \S\ref{sec:patching_parameters}.

\subsection{Time Complexity}
In order to create the patched indexes, the algorithm loops through the entire output file of both networks and tries to patch each graphlet against all graphlets within $P$ lines. If we denote $L$ as the number of lines in the larger file, the time complexity of this step is $O(LP)$. After creating patched indexes, the algorithm loops through the union of keys in both indexes and performs $O(1)$ work on each. To see this, notice how the algorithm checks the number of index entries for the key in both indexes and only outputs a patched graphlet (an $O(1)$ procedure) if both numbers are 1. Additionally, since the total number of keys is bounded by $2L$, the total time complexity is $O(LP+2L)$, which is just $O(LP)$.

\section{Experiments}
\label{sec:experiments}
In this section, we fix our parameters and measure the performance of our algorithm on a variety of real-world graphs from different domains. We discuss an issue we have identified regarding low volumes and explore possible improvements. Finally, we demonstrate the significance of our findings by comparing our seeds against seeds found from optimizing a common and effective measure of topological similarity.

All experiments are performed on a cluster of 96 identical machines (the ``circinus’’ cluster) in the Department of Computer Science at U.C. Irvine. Each host runs Linux CentOS, has 96GB of RAM and a 24-core Intel X5680 CPU running at 3.33GHz. Despite the large number of cores, the speed of a single core is comparable to that of a low-end laptop. BLANT-seed is part of the larger BLANT package \href{https://github.com/waynebhayes/BLANT}{available on Github}. The index creation algorithm (Algorithm \ref{alg:index}) is part of BLANT and is written in C, though the seed creation (Algorithm \ref{alg:seed}) is in Python.

As both our algorithm and the baseline method are deterministic, we perform one trial for each experiment.

\subsection{Datasets}
\label{sec:datasets}
We evaluate our algorithm on three different groups of networks, listed in \autoref{tab:networks}. First, we use the protein-protein interaction (PPI) networks of yeast from \cite{Collins2007yeast2}, which is a classic dataset amongst network alignment algorithms. This dataset consists of six networks which each allow edges (interactions) from an experiment at different confidence levels, with the percentage in the name (0 to 25\% in 5\% increments) indicate what fraction from the lower-confidence edge set is included.

Next, we use the rat and mouse PPI networks from the Integrated Interactions Database \cite{kotlyar2018iid}. The IID networks are, by far, the largest PPI networks available. Although they are partly synthetic, they are currently the best available approximation to ``real'' PPI networks that (a) are nontrivial in size, and (b) share what is believed to be about the same amount of topological similarity as we expect in the (currently unknown) true networks.

Finally, we use all but two of the temporal networks from the Stanford Network Analysis Project database, {\it SNAP} \cite{snapnets,kumar2018community,temporalNetMotifs,patternsDynamicsOnline,kumar2016edge,kumar2018rev2}. We ignore CommResistance because it is too small (fewer than 10 nodes per network). We ignore ActMOOC because they have two different types of nodes: users and activities.
Each temporal network consists of a list of edges, each with a timestamp. To generate a network, we collect all edges between a start time and an end time, in what we call a ``temporal window''. We create four windows total, and each window contains the same number of edges\footnote{We determine this fixed number of edges by starting from the beginning of the temporal network and adding edges one by one until we either hit 20,000 nodes, 400,000 edges, an edge/node ratio of 20:1, or we run out of edges. We created windows based on nodes/edges instead of times because the distribution of edges over time was far from linear in many networks. This allows us to create networks at comparable sizes and densities as the IID networks.}. The first window begins with the earliest edge in the network. The second window begins by shifting the start time until we have ``lost'' 1\% of the edges. The third and fourth windows use 3\% and 5\% shifting. The end times are determined by collecting edges until we have hit the set amount.

\begin{table}[htbp]
\caption{Network Statistics}
\label{tab:networks}
\begin{center}
\begin{tabular}{|c|c|c|}
    \hline
    \textbf{Networks} & \textbf{\# Nodes} & \textbf{\# Edges} \\
    \hline
    Yeast 0\%-25\% & 1.0K & 8.3K-10K \\
    \hline
    All IIDs & 13K-18K & 256K-335K \\
    \hline
    AskUbuntu 0\%-5\% & 20K & 49K\\
    \hline
    BitcoinAlpha 0\%-5\% & 3.5K & 13K \\
    \hline
    BitcoinOTC 0\%-5\% & 5.5K & 19K \\
    \hline
    CollegeMsg 0\%-5\% & 1.0K & 5.5K \\
    \hline
    EmailEUcore 0\%-5\% & 682 & 2.9K \\
    \hline
    MathOverflow 0\%-5\% & 20K & 82K \\
    \hline
    RedditHyperlinks 0\%-5\% & 20K & 60K \\
    \hline
    StackOverflow 0\%-5\% & 20K & 191K \\
    \hline
    SuperUser 0\%-5\% & 20K & 80K \\
    \hline
    WikiTalk 0\%-5\% & 20K & 82K \\
    \hline
\end{tabular}
\end{center}
\end{table}

\subsection{Baselines}
To the best of our knowledge, there is no direct competitor to BLANT-seed + BLANT-extend; together, they constitute the first local network alignment algorithm that works well using topology alone. We believe the most appropriate comparison follows Meng {\it et al.,}\cite{meng2016local}, who provided AlignMCL \cite{alignMCL} with a ranked list of all pairs of nodes in two networks sorted by {\it orbit degree vector (ODV)} similarity\footnote{This was described in Meng {\it et al.,}'s Supplementary Info. Note that what we refer to as an ODV---orbit degree vector---is often called a GDV---graphlet degree vector---in the literature, but in order to make the distinction clear between the contents of our graphlet and orbit degree vectors, BLANT uses these more descriptive names.}. While AlignMCL normally requires a list of {\em orthologs} as seeds, Meng {\it et al.,} substituted the top $n$ ODV-similarity pairs, where $n$ is the number of nodes in the smaller network. In this paper, we compare BLANT-seed to these topologically similar node pairs, which we will call {\it ODV seeds}. In BLANT-extend, we compare the BLANT-seed + BLANT-extend pipeline with the ODV-seeds + AlignMCL pipeline \cite{BLANT-extend}.

\subsection{Metrics}
\label{sec:metrics}
We measure the quality of local alignments using both volume and accuracy.

For volume, we measure both the raw number of unique $k$-graphlet seed pairs for $k$ from 8--15, as well as the volume of extracted node pairs (cf. Algorithm \ref{alg:seed}, \textit{function ExtractNodePairs}).

We use Node Correctness (NC, cf. \S\ref{sec:eval_metrics}) to measure accuracy of our extracted node pairs. For our graphlet pairs, we use a squared mean to capture the idea that ``1 seed with 8 correct pairs'' is better than ``8 seeds each with 1 correct pair''. We also weigh this sum by $k$, as larger graphlet seeds are more desirable. Equation \ref{eq:seed_nc} captures our ``weighted, squared mean'' method of measuring NC of a set of seeds; $t$ is the number of seeds, $s$ is the average size of a seed, $k_j$ is the size of the $j$th seed, and $c_j$ is the number of correct nodes in the $j$th seed.
    
\begin{equation}
    \mbox{Weighted Mean Graphlet NC} = \frac{1}{t * s} \sum_{j=1}^t {k_j (c_j ^ 2 / k_j ^ 2)}
    \label{eq:seed_nc}
\end{equation}

Finally, our topological accuracy is always 100\% for any edge-based measure such as EC \cite{GRAAL} or $S^3$ \cite{MAGNA} because---like BLAST's seeds that contain only perfect $k$-mer matches---our seed algorithm produces only perfect alignments between $k$-graphlets.

Since our goal is to find seeds, we value precision (accuracy) far more than recall (volume). This philosophy pervades the design of our algorithm as we only allow unambiguous graphlets, unique keys, and perfect topological matches. That said, a volume of zero would not be useful, and so we strive to increase volume only if it does not significantly harm accuracy. Thus, our full set of evaluation metrics is:

\begin{itemize}
    \item \textbf{Seed Volume (Seed Vol.)}: the number of seeds produced.
    \item \textbf{Mean Seed Size}: their average size.
    \item \textbf{Total Node Pairs}: total number of node pairs across all seed pairs. 
    \item \textbf{Extracted Volume (Extr. Vol.)}: the number of extracted node pairs.
    \item \textbf{Extracted Node Correctness (Extr. NC)}: fraction of extracted node pairs that are correct.
    \item \textbf{Seed Set Correctness}: weighted node correctness by (\ref{eq:seed_nc}).
    \item \textbf{Topological Similarity}: 100\% in all our cases, but useful for comparison with other methods.
\end{itemize}

\subsection{Fixed Parameters}
\label{sec:fixed_parameters}
The parameters $k$ and $D$ in Algorithm \ref{alg:index} were straightforward to select. As mentioned in \S\ref{sec:patch}, we want graphlets that are larger than $k=8$, so much so that we go through the effort of patching them together. Thus, we completely ignore generating $k=7$ and $k=6$ indexes. Selecting a value of $D$ was also easy: for smaller networks, we tried $D>2$ but $D=2$ was optimal; and on larger networks $D>2$ resulted in unacceptably long runtime (cf. \S\ref{sec:runtime}).

\subsection{Patching Parameters Experiment Setup}
\label{sec:patching_parameters}
In this section, we investigate the effects of the patching parameters $P$ (proximity in output file) and $C$ (minimum number of common nodes). We perform a factorial design experiment with both parameters and measure volume and accuracy. We slightly modify each parameter to better isolate the effects of individual values: instead of patching graphlets that are {\it within} $P$ lines and have {\it at least} $C$ nodes in common, we patch graphlets that are {\it exactly} $P'$ lines apart and have {\it exactly} $C'$ nodes in common.

We experimented with only a subset of networks to avoid over-fitting. For the IID networks, we chose mouse-rat and human-mouse as the two least synthetic pairs. Out of the five pairs of yeast networks, we chose yeast0-yeast05 because, as the pair with highest similarity, it maximizes the raw numbers in our results and reduces variance. Finally, for the temporal networks, we chose every other network in alphabetical order: AskUbuntu, BitcoinOTC, EmailEUCore, RedditHyperlinks, and SuperUser, using the 0\% and 1\% pair for each network.

\begin{figure}
    \centering
        \includegraphics[width=0.43\textwidth]{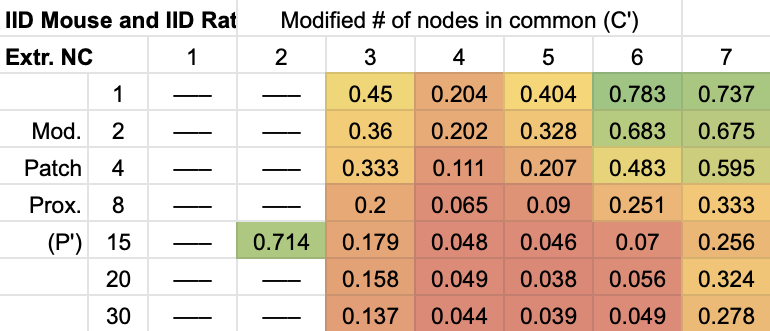}
        \includegraphics[width=0.43\textwidth]{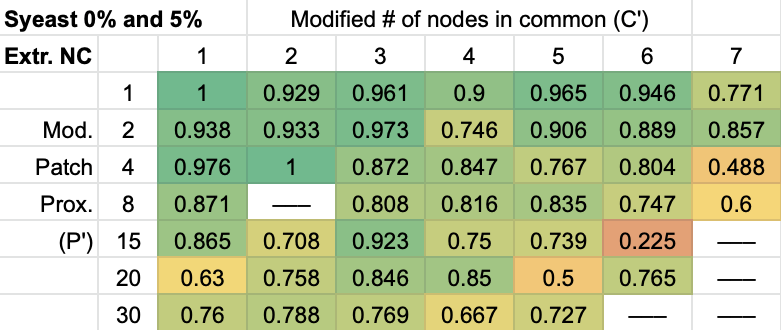}
        \includegraphics[width=0.43\textwidth]{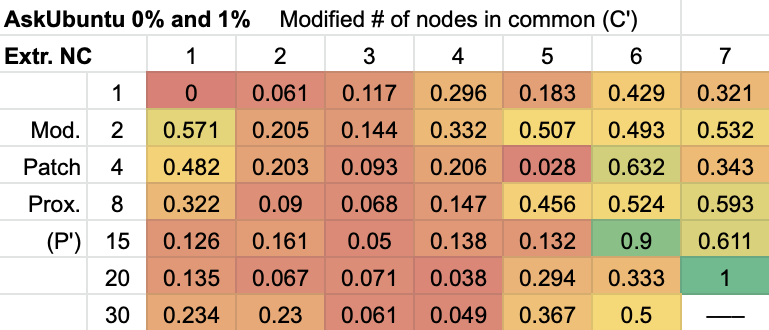}
        \includegraphics[width=0.43\textwidth]{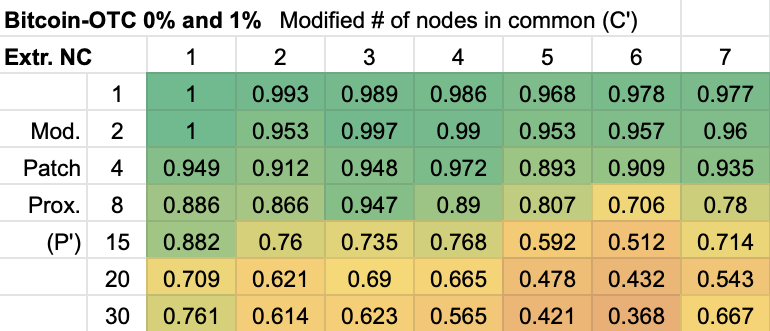}
        \includegraphics[width=0.43\textwidth]{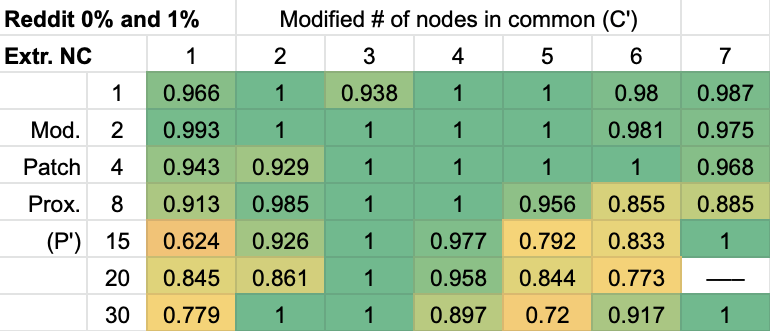}
        \includegraphics[width=0.43\textwidth]{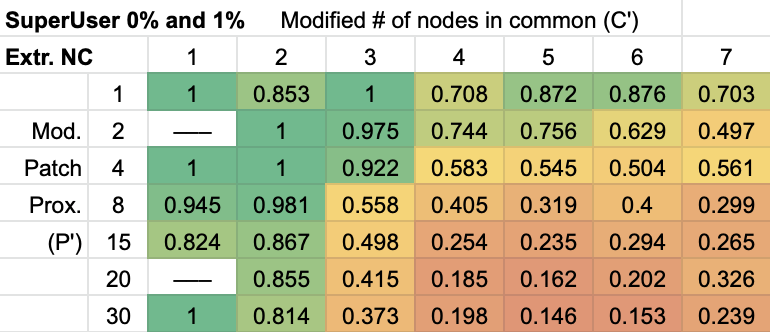}
    \caption{ Figures showing how Extracted Node Correctness (Extr. NC) varies over the $C'$ and $P'$ for our selected networks. We have ommitted ``IID human and mouse'' and ``EmailEUCore 0\% and 1\%'', as both had Extracted Volume counts in the single digits, making their Extr. NC values highly variable. (Red is bad, orange and yellow are OK, and green is good.)}
    \label{fig:patching}
\end{figure}

\begin{figure}
    \centering
        \includegraphics[width=0.43\textwidth]{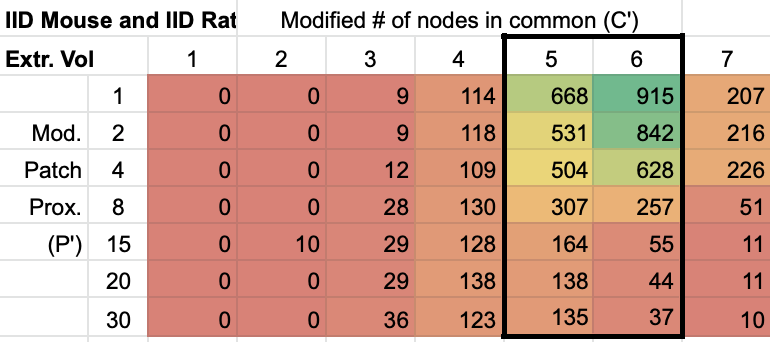}
        \includegraphics[width=0.43\textwidth]{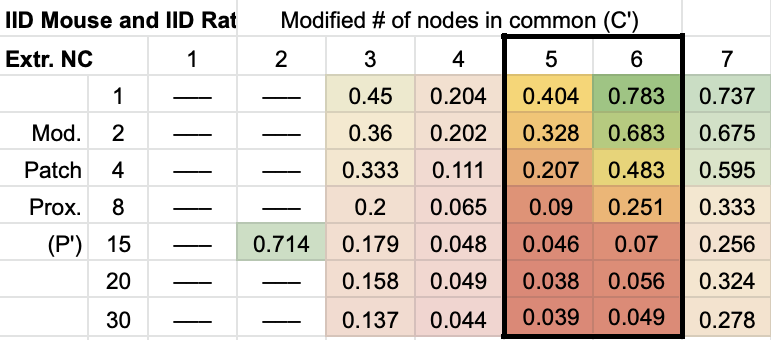}
        \includegraphics[width=0.43\textwidth]{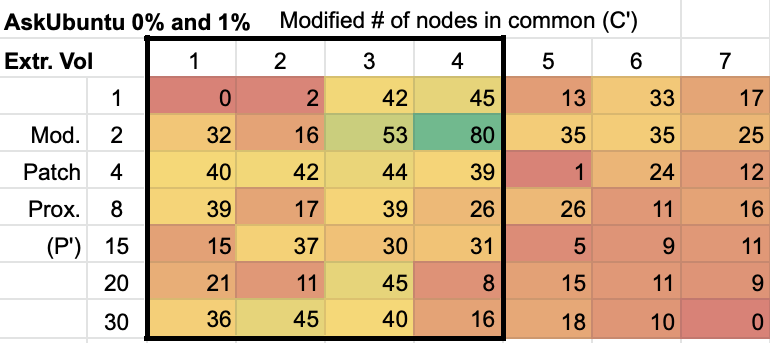}
        \includegraphics[width=0.43\textwidth]{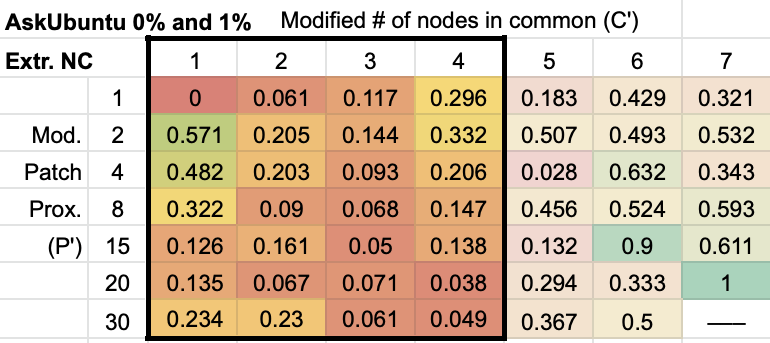}
        \includegraphics[width=0.43\textwidth]{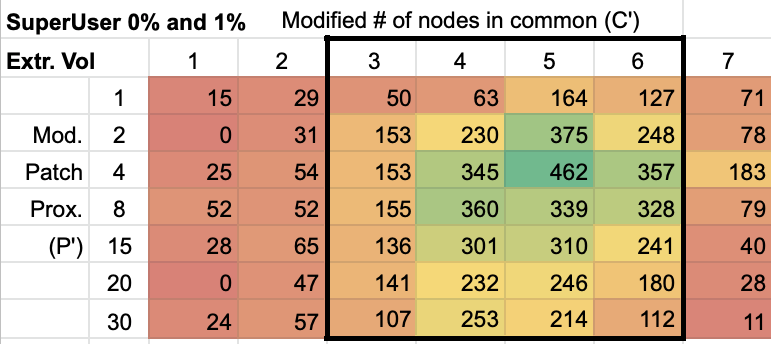}
        \includegraphics[width=0.43\textwidth]{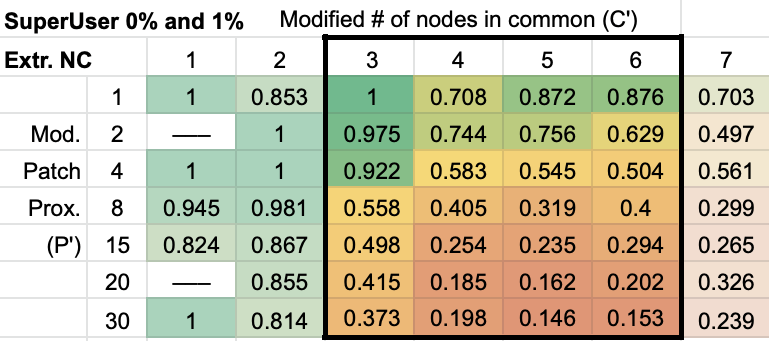}
    \caption{ Figures showing how Extracted Volume vary over the C' and P' for three network pairs. Interleaved are graphs highlighting the Extracted Node Correctness of these pairs in high volume regions.}
    \label{fig:patchzoom}
\end{figure}

As our main goal in creating seeds is achieving high accuracy, we mainly focus on how Extracted Node Correctness varies with these parameters in Fig.~\ref{fig:patching}. We only analyze how Extracted Volume changes to provide context when necessary, in Fig.~\ref{fig:patchzoom}.

\subsection{Patching Parameter Selections}
For the three network pairs in Fig.~\ref{fig:patching}, (Yeast 0\%, 5\%), (Bitcoin-OTC-0\%, 1\%), and (Reddit 0\%, 1\%), we can see that the accuracy decreases as $P'$ increases while the accuracy does not change much for different values of $C'$. Fig.~\ref{fig:patchzoom} shows that these same trends hold for (IID mouse, rat), (AskUbuntu 0\%, 1\%), and (SuperUser 0\%, 1\%) when we restrict our focus to higher volume regions of the network. In contrast, $P'$-$C'$ combinations which produce low volumes will naturally have a high variance in accuracy, and we can see a clearer trend by ignoring these low volume combinations (cf. Fig.~\ref{fig:patchzoom}).

Conceptually, the behaviour with respect to index file line difference makes sense: increasing $P'$ increases the distance between patched graphlets in the index file, which puts them further ``away'' from each other in our deterministic network traversal; if the networks have any topological differences at all, then each step in our algorithm has some chance of hitting such a ''difference'', so more steps means a greater chance of hitting differences, which destroys our ability to find the associated seeds. The tradeoff for $C'$ is less clear, since the effects are inconsistent across networks. A higher $C'$ value means more nodes in common, which means the graphlets were found in more similar contexts by the deterministic algorithm. However, a lower $C'$ value means a larger patched graphlet, making the shape more unique.

By analyzing the effects of $P'$ and $C'$, the best values of $C$ and $P$ may be selected. Accuracy is more important than volume when finding seeds, as a single high quality graphlet-seed appears sufficient for BLANT-extend to create a large, high quality alignment \cite{BLANT-extend}. Generally speaking, the Extr. NC does not change much as $C'$ changes when zooming in on high volume cells, so we select $C=1$---allowing any \# of nodes (except 0) to overlap.

$P$ can be adjusted based on the desired volume/accuracy tradeoff. As mentioned above, we care about accuracy more than volume, but we can't ignore volume entirely because more seeds gives more chances at one being suitable for extension. Thus, we choose $P=2$---corresponding to $P'\in \{1,2\}$---as a value which does not ignore volume completely but mostly focuses on high accuracy, especially in network pairs like IID mouse and rat, where accuracy decreases rapidly as $P'$ increases.

\subsection{Results}

\begin{figure}
    \centering
        \includegraphics[width=0.47\textwidth]{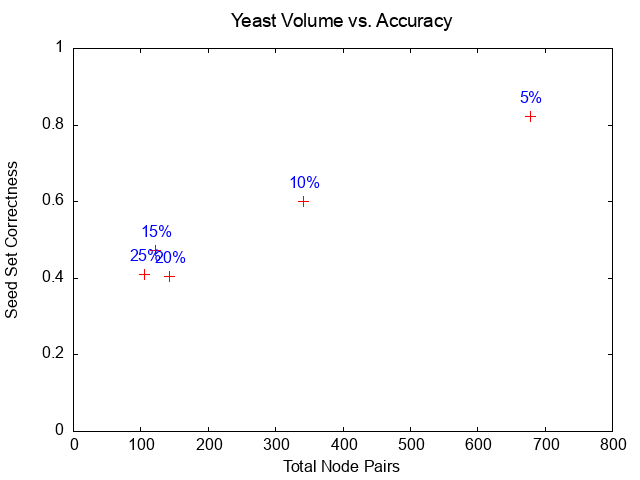}
        \includegraphics[width=0.47\textwidth]{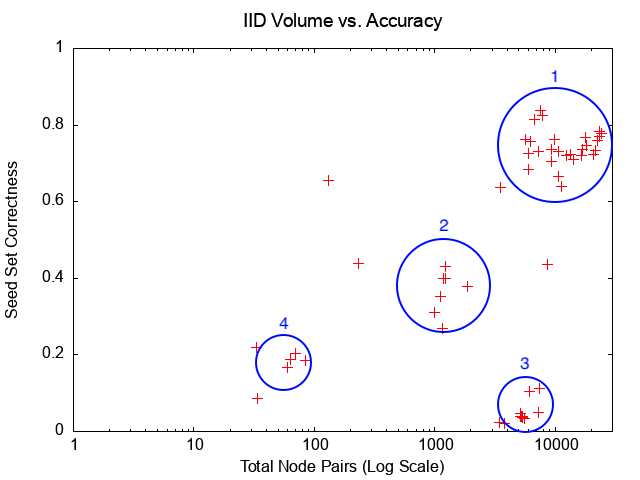}
        \includegraphics[width=0.47\textwidth]{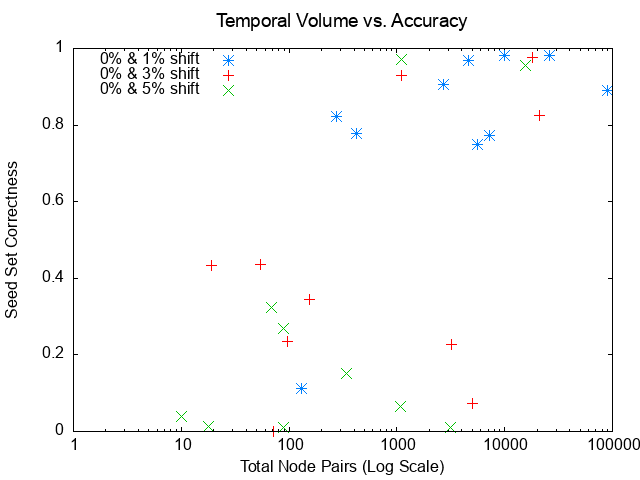}
    \caption{Accuracy {\it vs.} volume for Synthetic Yeast (top), IID (middle), and temporal networks (bottom). Note that IID and temporal volumes are log scale, yeast is linear. Yeast networks are labeled by the percentage of low-confidence edges being aligned with the ``all-confident'' (ie., 0\% low confidence) network. The temporal data is partitioned by shift level. The circles on the IID plot are ``clusters'' (see text).}
    \label{fig:volacc}
\end{figure}

Using the parameters described above, we tested BLANT-seed on all network pairs having the same type\footnote{We align yeast 0\% to 5\%-25\%. We align all 55 pairs of all 11 IID networks. We align each 0\% temporal network to the 1\%, 3\%, and 5\% networks, never aligning two different temporal networks.}. Fig.~\ref{fig:volacc} plots accuracy vs. volume with volume being measured by Total Node Pairs ($k=1$) and accuracy by Seed Set Correctness, for each of the three network types.

On all graphs, there tends to be a positive correlation between volume and accuracy. Combined with the labels in the yeast graph and the colored points in the temporal graph, we conclude that BLANT-seed achieves both higher volume and accuracy on networks with higher similarity. Even though the IID networks do not have a numerical similarity definition, we speculate that the volume and accuracy likely correlate with some true underlying measure of similarity.

Yeast shows the most stable results out of the three network types, as we did not need to graph its volume on a log scale. We achieve high accuracy and high volume (relative to the size of the network) for networks up to 25\% confidence level. This result is pleasing as yeast is one of the most common benchmarks in network alignment literature.

The IID networks seem to separate into four main clusters. In cluster 1, we achieve enormous volume (10K Total Node Pairs is about 1,000 $k\approx 10$ graphlet seeds) with high node correctness---about 0.8 on average. Cluster 2 shows fairly good volume (Total Node Pairs $\approx$ 1000) and fairly good accuracy (40\%, roughly on par with ODV seeds---discussed below---as our median is above their median, see \S\ref{sec:baseline_comparison}).

In cluster 3, we achieve a large volume but low accuracy. This is potentially a problem, but we observe that ODV seeds achieve low node correctness on many IID networks as well (cf. Fig.~\ref{fig:nc_box_plots}), demonstrating the difficulty of aligning some of these network pairs. Finally, cluster 4 shows both low volume and low accuracy. We are pleased that we achieve high performance on over half of network pairs, and improving the latter two clusters will require further study.

The temporal networks seemed to be grouped in two clusters in the top right and bottom left. Nearly all ``1\% shift'' points fall into the upper right and two 5\% networks still perform well. As with IID, we notice that ODV seeds also have low node correctness on many networks, demonstrating the difficulty of aligning these networks with the current state-of-the-art. As with IID, further study is warranted.

We omit figures of \textbf{topological similarity}, because our algorithm outputs only topologically perfect alignments by design.



\subsection{Initial Exploration into Sudden Volume Drops}
\label{sec:volume_drops}
\begin{figure}
    \centering
    \includegraphics[width=0.47\textwidth]{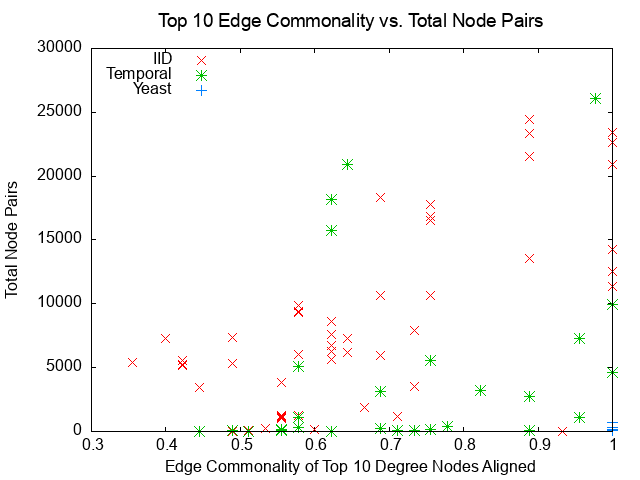}
    \caption{A graph showing the relationship between the similarity of hub nodes between two networks, and the volume of seeds. The top 10 nodes in two networks were each aligned in order (highest deg. to highest deg, 2nd highest to 2nd highest, etc.) and the edge commonality (EC) score of this alignment was computed. The EC score was 1.0 for all yeast networks, so the points are on the right border. }
    \label{fig:top10_vol}
\end{figure}

Of course, our algorithm is not without its weaknesses, and the main one is the sudden drop in volume and accuracy it experiences as networks become less similar, leading to the clusters we see in Fig.~\ref{fig:volacc}. We leave this problem to future study, but we have found some promising leads. First, Fig.~\ref{fig:top10_vol} shows that the volume is correlated with the EC (Edge Commonality, or fraction of aligned edges) of the 10 highest-degree nodes among the IID network pairs (but not temporal or yeast), which makes sense since our deterministic algorithm expands to neighbors of highest degree at each step. We hypothesize that this pattern holds among the IID networks because they have a very low diameter, causing nearly all expansions to reach the hub nodes within $k=8$ steps.

\begin{figure}
    \centering
        \includegraphics[width=0.47\textwidth]{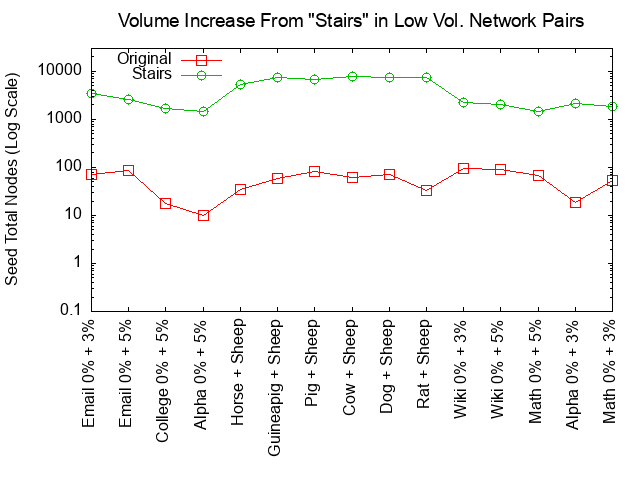}
        \includegraphics[width=0.47\textwidth]{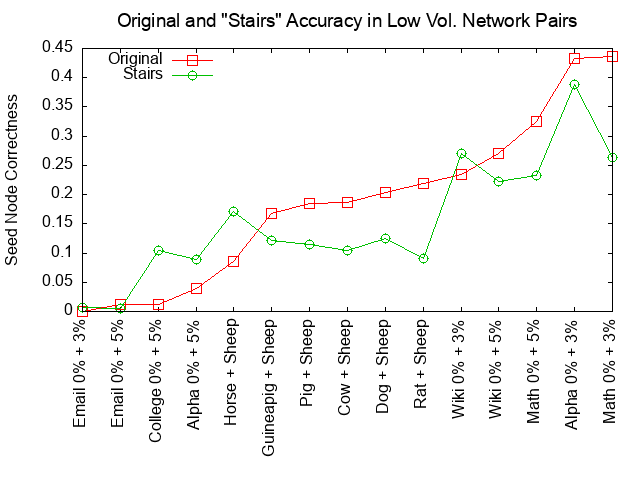}
    \caption{ We selected network pairs which produced $<100$ Total Node Pairs to investigate how to increase their volumes. The ``stairs'' algorithm produces significantly higher volumes (top figure, volume on a log scale) while mostly maintaining accuracy (bottom figure). Note that both figures use the same network ordering along the horizontal axis.}
    \label{fig:stairs_method}
\end{figure}

We tried many techniques in order to ``slow'' the algorithm from reaching hubs too quickly. One of these is the ``stairs'' algorithm, which simply skips the top $k-1-p$ degree values at each expansion step in Algorithm \ref{alg:index}, where $p$ is the number of nodes we currently have. We have found that this algorithm can achieve 10x more Total Node Pairs while maintaining accuracy on most networks with $<100$ Total Node Pairs (cf. Fig.~\ref{fig:stairs_method}, volume shown on log scale). This method performs worse as a whole on our dataset, however, so we have elected not to use it. While the $<100$ threshold is arbitrary, it shows that our algorithm has the potential to achieve our goal of $>1000$ Total Node Pairs on \textbf{any} type of network with less than 5\% shift.

\subsection{Comparison with Baselines}
\label{sec:baseline_comparison}
We use a pipeline of ODV seeds + AlignMCL for comparison with BLANT-seed + BLANT-extend. We will compare the first parts of these pipelines to each other in this paper, leaving the full comparison for our companion paper BLANT-extend. AlignMCL wants, as input, the set of orthologous pairs of nodes, which in the topological context converted to the top $n_{min}$ node pairs in ODV similarity ($n_{min}$ being the number of nodes in the smaller network) \cite{meng2016local}. Since \cite{meng2016local} used these parameters to conclude that AlignMCL was the best available topology-only local aligner, we adopt this list as well (though we offer a caveat in our Discussion below).

In addition to the work in \cite{meng2016local}, we found that ODV similarity alone provided poor results on large networks because it computed similarities of 1.0 for many leaf nodes (degree 1 nodes). We tried two methods to deal with this. First, we weigh node pairs with higher degree more with the $\alpha$ parameter described in \cite{GRAAL}'s Supplementary Info, using their recommended value of $\alpha=0.8$. Next, we present a new method which simply ignores all leaf nodes.

Before showing quantitative results, we list the major qualitative differences among these two seeding algorithms. First, our main technical achievement is that BLANT-seed outputs graphlet pairs, not just node pairs; our companion paper \cite{BLANT-extend} shows that graphlet seeds with $k>1$ are much more effectively extended than the $k=1$ seeds---{\it ie.,} node pairs. On top of this, our graphlet pairs have perfect topological similarity, a metric which does not exist for node pairs.

Seed volume is another qualitative difference. Since {\em every} pair of nodes has an ODV similarity, the number of ODV similarities from networks with $n_1$ and $n_2$ nodes is $n_1n_2$, which is {\em far} more than the number of orthologs---which must be less than $n_{min}=\min(n_1,n_2)$. Since there is no obvious cutoff threshold for ODV similarities, \cite{meng2016local} reasonably chose to use the top $n_{min}$ ODV similarities as ``orthologs''. We, on the other hand, strongly prefer accuracy in our graphlet seeds; the demand for high accuracy sometimes allows huge volumes, but in some circumstances severely restricts it.

We believe it is a {\em strength} that our mechanism is able to automatically restrict volume in an attempt to get higher accuracy; a sorted list of ODV similarities has no such mechanism, and so using the top $n_{min}$ of them seems the only fair comparison---though we discuss alternatives in our Discussion section.

With these qualitative differences in mind, the only axis on which we can compare the seeding components of these two pipelines is node correctness. To better represent our overall node correctness, we use Extr. NC rather than Seed NC as it outputs node pairs uniquely. As before, we partition our results by network type.

\begin{figure}
    \centering
        \includegraphics[width=0.47\textwidth]{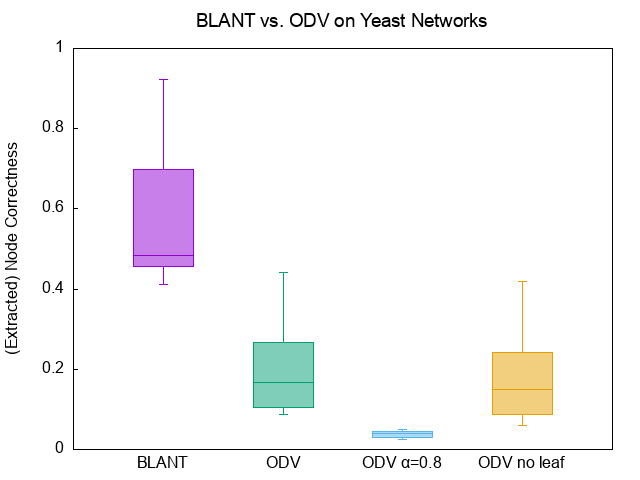}
        \includegraphics[width=0.47\textwidth]{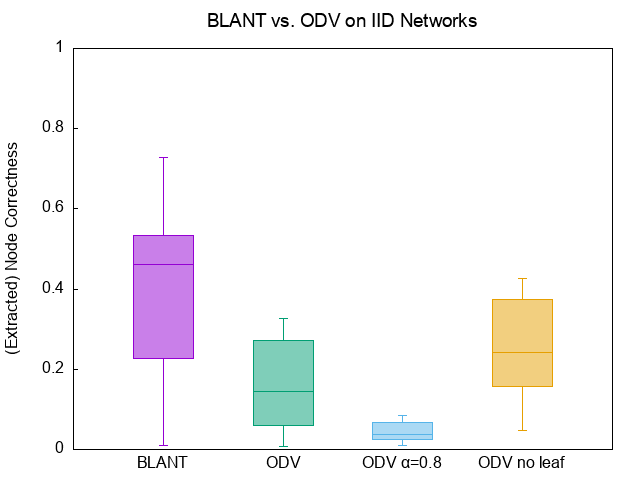}
        \includegraphics[width=0.47\textwidth]{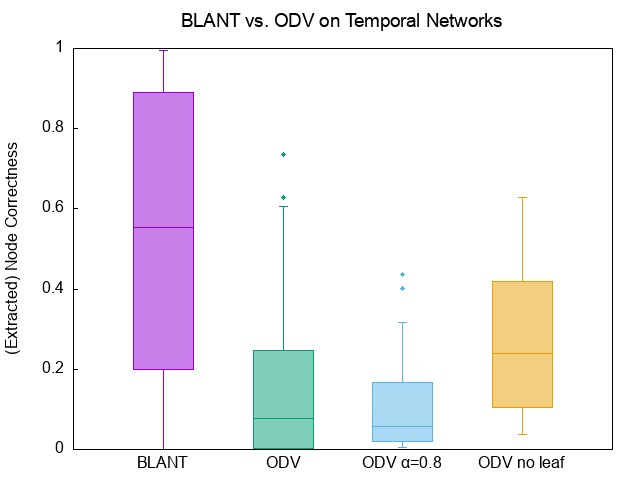}
    \caption{Box-and-whisker plots of Node Correctness for each ODV method (Extracted Node Correctness for BLANT), partitioned by network type, across all networks of that type. The box spans the 25th percentile to the 75th percentile, with the middle line being the median. The upper and lower whiskers show the highest and lowest node correctness values, with dots representing outliers.  BLANT-seed is compared against all three variations of ODV seeds: pure ODV similarity, using GRAAL's $\alpha=0.8$, and ignoring leaf nodes. }
    \label{fig:nc_box_plots}
\end{figure}

Fig.~\ref{fig:nc_box_plots} shows that our node correctness significantly outperforms ODV seeds on all networks, even after we improved the ODV method as described above. While this does not demonstrate superiority over ODV seeds due to the above qualitative differences, it does show that we achieve strong node correctness values, on top of our perfect topological similarity. 

\subsection{Runtime}
\label{sec:runtime}
With runtime, our goal is to be able to run the algorithm in a reasonable amount of time on a standard laptop. This measure is fuzzy as runtime is not a primary concern of ours, but we speculate that a researcher (the likely user of our algorithm) would feel that ``a few hours'' is a reasonable one-time cost to index a network. We are happy to report that our index creation runtimes fit comfortably in this range, as shown in \autoref{tab:runtimes}. After creating the index, the researcher only needs to run our linear-time seed mining algorithm, which takes less than a minute on all network pairs despite being written in Python.

Additionally, we note that our algorithm may be easily parallelized to greatly reduce runtime.

\begin{table}[htbp]
\caption{Runtime Statistics Across All Networks}
\label{tab:runtimes}
\begin{center}
\begin{tabular}{|c|c|c|c|c|}
    \hline
    \textbf{Units} & \textbf{Minimum} & \textbf{Median} & \textbf{Max} & \textbf{Std. Deviation} \\
    \hline
    Minutes & 0.055 & 32.23 & 58.46 & 16.53 \\
    \hline
\end{tabular}
\end{center}
\end{table}

We do not rigorously compare our runtime with that of creating ODV seeds, since runtime is not a primary concern of our algorithm and since we wrote the ODV seed code in Python. We only note that the Python ODV seed code takes 1-4 hours to generate results for each network pair.

\section{Discussion}
To the best of our knowledge, we are the first to present a topology-only seed-identification algorithm which discovers pairs of \textit{graphlets}, not just pairs of nodes. On top of that, we have demonstrated that our topologically perfect alignments of up to $k=15$ nodes also have superior node correctness values compared to ODV node pairs, a popular topology-only seeding method. While our algorithm does have inconsistent volume, we have discovered a few promising directions to solve this issue.

We have achieved these seeding results by utilizing a key innovation: a deterministically generated graphlet index of the network. The use of determinism is crucial as it exploits the actual similarity---if present---among the set of networks in a way that exponentially reduces the search space.

The NC comparison presented in Fig.~\ref{fig:nc_box_plots} depends strongly on the volume of the two methods. While our volume is not determined beforehand but is instead a consequence of our algorithm, the volume of ODV similarities is predetermined at $n_{min}$---the number of nodes in the smaller network. In some cases, our volume---call it $\nu$---is much smaller than $n_{min}$; in this case one could argue that we should look at the NC of the top-$\nu$ ODV similarities. Predictably, the NC of a list of sorted ODV similarities increases closer to the top, and so if we use {\em our} volume as the number of top-ODV-similarities to take, the resulting NC values are comparable---and in some cases exceed---ours. However, in the absence of an independent method of choosing a threshold on ODV similarities---and since AlignMCL demands a list of $n_{min}$ pairs---we believe the fairest comparison is to use our volume for our NC, and $n_{min}$ for the ODV similarity set.

Finally, our seeds have also been validated ``in the field'' in that BLANT-extend can discover 280-node, topologically perfect alignments between the two large networks of IID mouse and rat that have high Resnik semantic similarity.

\section*{Acknowledgment}
We thank Brian Song and Ronit Barman for assisting with data collection and graph generation, and Arthur Jiejie Lafrance for implementing the heuristic function in the index creation algorithm.


\bibliographystyle{IEEEtran}
\bibliography{wayne-all}

\begin{thebibliography}{10}
\providecommand{\url}[1]{#1}
\csname url@samestyle\endcsname
\providecommand{\newblock}{\relax}
\providecommand{\bibinfo}[2]{#2}
\providecommand{\BIBentrySTDinterwordspacing}{\spaceskip=0pt\relax}
\providecommand{\BIBentryALTinterwordstretchfactor}{4}
\providecommand{\BIBentryALTinterwordspacing}{\spaceskip=\fontdimen2\font plus
\BIBentryALTinterwordstretchfactor\fontdimen3\font minus
  \fontdimen4\font\relax}
\providecommand{\BIBforeignlanguage}[2]{{%
\expandafter\ifx\csname l@#1\endcsname\relax
\typeout{** WARNING: IEEEtran.bst: No hyphenation pattern has been}%
\typeout{** loaded for the language `#1'. Using the pattern for}%
\typeout{** the default language instead.}%
\else
\language=\csname l@#1\endcsname
\fi
#2}}
\providecommand{\BIBdecl}{\relax}
\BIBdecl

\bibitem{emmert2016fifty}
F.~Emmert-Streib, M.~Dehmer, and Y.~Shi, ``Fifty years of graph matching,
  network alignment and network comparison,'' \emph{Information Sciences}, vol.
  346, pp. 180--197, 2016.

\bibitem{clark2014comparison}
C.~Clark and J.~Kalita, ``A comparison of algorithms for the pairwise alignment
  of biological networks,'' \emph{Bioinformatics}, vol.~30, no.~16, pp.
  2351--2359, 2014.

\bibitem{Dehmer2011}
\BIBentryALTinterwordspacing
M.~Dehmer and F.~Emmert-Streib, \emph{Mining Graph Patterns in Web-based
  Systems: A Conceptual View}.\hskip 1em plus 0.5em minus 0.4em\relax
  Dordrecht: Springer Netherlands, 2011, pp. 237--253. [Online]. Available:
  \url{https://doi.org/10.1007/978-90-481-9178-9_11}
\BIBentrySTDinterwordspacing

\bibitem{Sommerfeld1994}
\BIBentryALTinterwordspacing
E.~Sommerfeld and F.~Sobik, \emph{Operations on cognitive structures --- their
  modeling on the basis of graph theory}.\hskip 1em plus 0.5em minus
  0.4em\relax Berlin, Heidelberg: Springer Berlin Heidelberg, 1994, pp.
  151--196. [Online]. Available:
  \url{https://doi.org/10.1007/978-3-642-52064-8_5}
\BIBentrySTDinterwordspacing

\bibitem{metaDiagramAlignment}
Y.~Ren, C.~C. Aggarwal, and J.~Zhang, ``Meta diagram based active social
  networks alignment,'' in \emph{2019 IEEE 35th International Conference on
  Data Engineering (ICDE)}, 2019, pp. 1690--1693.

\bibitem{HSIEH2008401}
\BIBentryALTinterwordspacing
S.-M. Hsieh and C.-C. Hsu, ``Graph-based representation for similarity
  retrieval of symbolic images,'' \emph{Data \& Knowledge Engineering},
  vol.~65, no.~3, pp. 401--418, 2008. [Online]. Available:
  \url{https://www.sciencedirect.com/science/article/pii/S0169023X07002212}
\BIBentrySTDinterwordspacing

\bibitem{GareyJohnson}
M.~Garey and D.~Johnson, \emph{Computers and Intractability: A Guide to the
  Theory of NP-Completeness}.\hskip 1em plus 0.5em minus 0.4em\relax New York:
  New York: W.H. Freeman, 1979.

\bibitem{meng2016local}
L.~Meng, A.~Striegel, and T.~Milenkovi{\'c}, ``Local versus global biological
  network alignment,'' \emph{Bioinformatics}, vol.~32, no.~20, pp. 3155--3164,
  2016.

\bibitem{LIU2018318}
\BIBentryALTinterwordspacing
L.~Liu, B.~Qu, B.~Chen, A.~Hanjalic, and H.~Wang, ``Modelling of information
  diffusion on social networks with applications to wechat,'' \emph{Physica A:
  Statistical Mechanics and its Applications}, vol. 496, pp. 318--329, 2018.
  [Online]. Available:
  \url{https://www.sciencedirect.com/science/article/pii/S0378437117312785}
\BIBentrySTDinterwordspacing

\bibitem{alignMCL}
M.~Mina and P.~Guzzi, ``Alignmcl: Comparative analysis of protein interaction
  networks through markov clustering,'' in \emph{Proceedings - 2012 IEEE
  International Conference on Bioinformatics and Biomedicine Workshops, BIBMW
  2012}, 10 2012, pp. 174--181.

\bibitem{MamanoHayesSANA}
N.~Mamano and W.~B. Hayes, ``{SANA}: Simulated annealing far outperforms many
  other search algorithms for biological network alignment.''
  \emph{Bioinformatics (Oxford, England)}, vol.~33, pp. 2156--2164, 2017.

\bibitem{regal2018}
\BIBentryALTinterwordspacing
M.~Heimann, H.~Shen, T.~Safavi, and D.~Koutra, ``Regal,'' \emph{Proceedings of
  the 27th ACM International Conference on Information and Knowledge
  Management}, Oct 2018. [Online]. Available:
  \url{http://dx.doi.org/10.1145/3269206.3271788}
\BIBentrySTDinterwordspacing

\bibitem{dana2019}
\BIBentryALTinterwordspacing
T.~Derr, H.~Karimi, X.~Liu, J.~Xu, and J.~Tang, ``Deep adversarial network
  alignment,'' \emph{CoRR}, vol. abs/1902.10307, 2019. [Online]. Available:
  \url{http://arxiv.org/abs/1902.10307}
\BIBentrySTDinterwordspacing

\bibitem{wang2022sana}
S.~Wang, G.~R.~S. Atkinson, and W.~B. Hayes, ``Sana: Cross-species prediction
  of gene ontology go annotations via topological network alignment,'' 2022.

\bibitem{WeBeat}
S.~Wang, X.~Chen, B.~J. Frederisy, B.~A. Mbakogu, A.~D. Kanne, P.~Khosravi, and
  W.~B. Hayes, ``On the current failure---but bright future---of
  topology-driven biological network alignment,'' \emph{{Advances in Protein
  Chemistry and Structural Biology (accepted; preprint
  https://doi.org/10.48550/arXiv.2204.11999)}}, 2022.

\bibitem{GRAAL}
O.~Kuchaiev, T.~Milenkovi{\'c}, V.~Memi{\v s}evi{\'c}, W.~Hayes, and
  N.~Pr\v{z}ulj, ``Topological network alignment uncovers biological function
  and phylogeny,'' \emph{Journal of The Royal Society Interface}, vol.~7,
  no.~50, pp. 1341--1354, 2010.

\bibitem{Tijana2008}
T.~Milenkovi\'{c} and N.~Pr\v{z}ulj, ``Uncovering biological network function
  via graphlet degree signatures,'' \emph{Cancer Inform.}, vol.~6, no. Epub
  2008 Apr 14, pp. 257--273, 2008.

\bibitem{BLANT-extend}
T.~Ding, U.~Jain, and W.~B. Hayes, ``{BLANT: Basic Local Alignment of Network
  Topology, Part 2: Topology-only Extension Beyond Graphlet Seeds},''
  \emph{submitted}, 06 2022.

\bibitem{blanttool}
\BIBentryALTinterwordspacing
S.~Maharaj, B.~Tracy, and W.~B. Hayes, ``{BLANT - Fast Graphlet Sampling
  Tool},'' \emph{Bioinformatics}, 08 2019. [Online]. Available:
  \url{https://doi.org/10.1093/bioinformatics/btz603}
\BIBentrySTDinterwordspacing

\bibitem{hayesblant}
W.~Hayes and S.~Maharaj, ``{BLANT: Sampling Graphlets in a Flash},'' in
  \emph{q-bio}, 2018.

\bibitem{ORCA}
\BIBentryALTinterwordspacing
T.~Ho\v{c}evar and J.~Dem\v{s}ar, ``{A combinatorial approach to graphlet
  counting},'' \emph{Bioinformatics}, vol.~30, no.~4, pp. 559--565, Feb. 2014.
  [Online]. Available: \url{http://dx.doi.org/10.1093/bioinformatics/btt717}
\BIBentrySTDinterwordspacing

\bibitem{Melckenbeeck1EtAl2017}
I.~Melckenbeeck, P.~Audenaert, D.~Colle, and M.~Pickavet, ``Efficiently
  counting all orbits of graphlets of any order in a graph using autogenerated
  equations,'' \emph{preprint}, 2017.

\bibitem{Altschul90}
S.~F. Altschul, W.~Gish, W.~Miller, and D.~J. Lipman, ``Basic local alignment
  search tool,'' \emph{Journal of Molecular Biology}, vol. 215, pp. 403--410,
  1990.

\bibitem{Przulj2004Graphlets}
\BIBentryALTinterwordspacing
N.~Pr\v{z}ulj, D.~G. Corneil, and I.~Jurisica, ``Modeling interactome:
  scale-free or geometric?'' \emph{Bioinformatics}, vol.~20, no.~18, pp.
  3508--3515, 2004. [Online]. Available:
  \url{http://bioinformatics.oxfordjournals.org/content/20/18/3508.abstract}
\BIBentrySTDinterwordspacing

\bibitem{Melckenbeeck1EtAl2016}
I.~Melckenbeeck, P.~Audenaert, T.~Michoel, D.~Colle, and M.~Pickavet, ``An
  algorithm to automatically generate the combinatorial orbit counting
  equations,'' \emph{PLoS ONE}, vol.~11, no.~1, 2016.

\bibitem{melckenbeeck2017efficiently}
I.~Melckenbeeck, P.~Audenaert, D.~Colle, and M.~Pickavet, ``Efficiently
  counting all orbits of graphlets of any order in a graph using autogenerated
  equations,'' \emph{Bioinformatics}, vol.~1, p.~9, 2017.

\bibitem{MAGNA}
V.~Saraph and T.~Milenkovi{\'c}, ``{MAGNA}: maximizing accuracy in global
  network alignment,'' \emph{Bioinformatics}, vol.~30, no.~20, pp. 2931--2940,
  2014.

\bibitem{hasan2017graphettes}
A.~Hasan, P.-C. Chung, and W.~Hayes, ``Graphettes: Constant-time determination
  of graphlet and orbit identity including (possibly disconnected) graphlets up
  to size 8,'' \emph{PloS one}, vol.~12, no.~8, p. e0181570, 2017.

\bibitem{NAUTY}
\BIBentryALTinterwordspacing
B.~D. Mckay, ``Nauty,'' 2010. [Online]. Available:
  \url{http://users.cecs.anu.edu.au/~bdm/nauty}
\BIBentrySTDinterwordspacing

\bibitem{Collins2007yeast2}
\BIBentryALTinterwordspacing
S.~R. Collins, P.~Kemmeren, X.-C. Zhao, J.~F. Greenblatt, F.~Spencer, F.~C.~P.
  Holstege, J.~S. Weissman, and N.~J. Krogan, ``Toward a comprehensive atlas of
  the physical interactome of saccharomyces cerevisiae,'' \emph{Molecular and
  Cellular Proteomics}, vol.~6, no.~3, pp. 439--450, 2007. [Online]. Available:
  \url{http://www.mcponline.org/content/6/3/439.abstract}
\BIBentrySTDinterwordspacing

\bibitem{kotlyar2018iid}
M.~Kotlyar, C.~Pastrello, Z.~Malik, and I.~Jurisica, ``{IID} 2018 update:
  context-specific physical protein--protein interactions in human, model
  organisms and domesticated species,'' \emph{Nucleic acids research}, vol.~47,
  no.~D1, pp. D581--D589, 2018.

\bibitem{snapnets}
J.~Leskovec and A.~Krevl, ``{SNAP Datasets}: {Stanford} large network dataset
  collection,'' \url{http://snap.stanford.edu/data}, Jun. 2014.

\bibitem{kumar2018community}
S.~Kumar, W.~L. Hamilton, J.~Leskovec, and D.~Jurafsky, ``Community interaction
  and conflict on the web,'' in \emph{Proceedings of the 2018 World Wide Web
  Conference on World Wide Web}.\hskip 1em plus 0.5em minus 0.4em\relax
  International World Wide Web Conferences Steering Committee, 2018, pp.
  933--943.

\bibitem{temporalNetMotifs}
\BIBentryALTinterwordspacing
A.~Paranjape, A.~R. Benson, and J.~Leskovec, ``Motifs in temporal networks,''
  in \emph{Proceedings of the Tenth ACM International Conference on Web Search
  and Data Mining}, ser. WSDM '17.\hskip 1em plus 0.5em minus 0.4em\relax New
  York, NY, USA: Association for Computing Machinery, 2017, p. 601–610.
  [Online]. Available: \url{https://doi.org/10.1145/3018661.3018731}
\BIBentrySTDinterwordspacing

\bibitem{patternsDynamicsOnline}
P.~Panzarasa, T.~Opsahl, and K.~Carley, ``Patterns and dynamics of users'
  behavior and interaction: Network analysis of an online community,''
  \emph{JASIST}, vol.~60, pp. 911--932, 05 2009.

\bibitem{kumar2016edge}
S.~Kumar, F.~Spezzano, V.~Subrahmanian, and C.~Faloutsos, ``Edge weight
  prediction in weighted signed networks,'' in \emph{Data Mining (ICDM), 2016
  IEEE 16th International Conference on}.\hskip 1em plus 0.5em minus
  0.4em\relax IEEE, 2016, pp. 221--230.

\bibitem{kumar2018rev2}
S.~Kumar, B.~Hooi, D.~Makhija, M.~Kumar, C.~Faloutsos, and V.~Subrahmanian,
  ``Rev2: Fraudulent user prediction in rating platforms,'' in
  \emph{Proceedings of the Eleventh ACM International Conference on Web Search
  and Data Mining}.\hskip 1em plus 0.5em minus 0.4em\relax ACM, 2018, pp.
  333--341.

\end{thebibliography}
\end{document}